\documentclass[12pt]{article}
\usepackage{amsmath}
\usepackage{amssymb}

\input{tcilatex}
\begin{document}

\begin{center}
\textbf{ALTERNATIVE SELF-DUAL GRAVITY}

\smallskip \ 

\textbf{IN EIGHT DIMENSIONS}

\smallskip \ 

\smallskip \ 

\smallskip \ 

J. A. Nieto\footnote{%
niet@uas.edu.mx; janieto1@asu.edu}

\smallskip \ 

\textit{Facultad de Ciencias F\'{\i}sico-Matem\'{a}ticas de la Universidad
Aut\'{o}noma} \textit{de Sinaloa, 80010, Culiac\'{a}n Sinaloa, M\'{e}xico.}

\bigskip \ 

\bigskip \ 

\textbf{Abstract}
\end{center}

We develop an alternative Ashtekar formalism in eight dimensions. In fact,
using a MacDowell-Mansouri physical framework and a self-dual curvature
symmetry we propose an action in eight dimensions in which the Levi-Civita
tenor with eight indices plays a key role. We explicitly show that such an
action contains number of linear, quadratic and cubic terms in the Riemann
tensor, Ricci tensor and scalar curvature. In particular, the linear term is
reduced to the Einstein-Hilbert action with cosmological constant in eight
dimensions. We prove that such a reduced action is equivalent to the
Lovelock action in eight dimensions.

\bigskip \ 

\bigskip \ 

\bigskip \ 

\bigskip \ 

\bigskip \ 


\noindent Pacs numbers: 04.60.-m, 04.65.+e, 11.15.-q, 11.30.Ly

\noindent 29, July 2016

\newpage

\noindent \textbf{1.- Introduction}

\smallskip \ 

Higher dimensional Ashtekar formalism has shown to be a promising proposal
[1-5] (see also Ref. [6-10]). The main physical reason for this is that it
allows a possible connection with string theory [11]; an alternative
approach for quantum gravity. However, except for the $8$-dimensional
Ashtekar formalism [6] the concept so important in $4$-dimensions of the
self duality curvature concept is lost. Such an $8$-dimensional Ashtekar
formalism is an interesting approach based on the octonionic structure.
Here, we would like to present an alternative approach in $8$-dimensional
containing the self-duality curvature concept that is even closer to one
considered in $4$-dimensions.

Another source of physical interest in the present work it can be found from
the fact that versions of $8$-dimensional theory, such as $(4+4)$%
-dimensional scenario, may be obtained from dimensional reduction of a $%
(5+5) $-dimensional theory which is one of the possible background
candidates for the so called $M$-theory (see Refs. [12]-[15]). In fact, the $%
(4+4)$-dimensional structure emerges from Majorana-Weyl constraints applied
to superstrings [11] and supergravity [16]. In this context, it has been
shown that the triality automorphisms of $Spin(8)$ act on Majorana-Weyl
representations leading to relations among $(1+9)\leftrightarrow
(5+5)\leftrightarrow (9+1)$ signatures, as well as their corresponding
transverse signatures $(0+8)\leftrightarrow (4+4)\leftrightarrow (8+0)$
[13]. Finally, it has been shown that the $(4+4)$-dimensional theory has an
interesting connection with qubits and chirotopes (see Refs. [17]-[21] and
references therein). So, one may expect that eventually the $(4+4)$%
-dimensional Ashtekar formalism may shed some light on $M$-theory.

In this work, it is considered a self (antiself)-dual theory \textit{a la}
Ashtekar in $8$-dimensions. Specifically, using the MacDowell-Mansouri
physical framework and a self-dual curvature symmetry we propose an action
in $8$-dimensions in which the the $\epsilon $-symbol (Levi-Civita density
tensor) with $8$-indices $\epsilon ^{a_{1}...a_{8}}$ plays a key role. We
explicitly show that such an action contains the Einstein-Hilbert action
with cosmological constant in $8$-dimensions. In addition, it also contains
a number of quadratic and cubic terms in the Riemann tensor, Ricci tensor
and scalar curvature. In fact, we prove that such a reduced action is
equivalent to the Lovelock action in $8$-dimensions.

The plan of this work is the following. In section 2, we briefly recall the
self (antiself)-duality formalism in $4$-dimensions. In section 3 and 4, we
introduce the analogue formalism in $8$-dimensions and prove that such a
formalism is equivalent to the Lovelock theory in $8$-dimensions. Finally,
in section 5, we summarize our results and make some final comments.

\smallskip \ 

\noindent \textbf{2.- Ashtekar formalism in 4-dimensions}

\smallskip \ 

Let us start considering a metric $g_{\mu \nu }=e_{\mu }^{a}e_{\nu }^{b}\eta
_{ab}$ in terms of the vielbien $e_{\mu }^{a}$ and a flat metric $\eta _{ab}$%
. In addition let us introduce the MacDowell-Mansouri kind of curvature
tensor in 4-dimensions:

\begin{equation}
\mathcal{R}_{\mu \nu }^{ab}=R_{\mu \nu }^{ab}+\lambda \Sigma _{\mu \nu
}^{ab}.  \label{1}
\end{equation}%
Here, $R_{\mu \nu }^{ab}$ is a curvature tensor defined, in the usual way,
in terms of a $SO(1,3)$-connection $\omega _{\mu }^{ab}$, namely 
\begin{equation}
R_{\mu \nu }^{ab}=\partial _{\mu }\omega _{\nu }^{ab}-\partial _{\nu }\omega
_{\mu }^{ab}+\partial \omega _{\mu }^{ac}\partial \omega _{\nu }^{db}\eta
_{cd}-\partial \omega _{\nu }^{ac}\partial \omega _{\mu }^{db}\eta _{cd},
\label{2}
\end{equation}%
while

\begin{equation}
\Sigma _{\mu \nu }^{ab}=e_{\mu }^{a}e_{\nu }^{b}-e_{\mu }^{b}e_{\nu }^{a}.
\label{3}
\end{equation}%
Moreover, $\lambda $ is a constant parameter which can be related with the
cosmological constant. In fact, this parameter can be properly reabsorbed in 
$e_{\mu }^{a}$ in such way that, for the purpose of computations, one can
set it as $\lambda =1$, but eventually it can be recovered in the final
result.

In 1996, Nieto, Socorro and Obregon [22] proposed the self (antiself)-dual
action

\begin{equation}
S=\frac{1}{2^{2}}\int d^{4}x\epsilon ^{\mu _{1}\mu _{2}\mu _{3}\mu
_{4}}\epsilon _{a_{1}a_{2}a_{3}a_{4}}\text{ }^{\pm }\mathcal{R}_{\mu _{1}\mu
_{2}}^{a_{1}a_{2}}\text{ }^{\pm }\mathcal{R}_{\mu _{3}\mu _{4}}^{a_{3}a_{4}}.
\label{4}
\end{equation}%
Here,

\begin{equation}
^{\pm }\mathcal{R}_{\mu _{1}\mu _{2}}^{a_{1}a_{2}}=\frac{1}{2}(\mathcal{R}%
_{\mu _{1}\mu _{2}}^{a_{1}a_{2}}+\xi \epsilon _{a_{3}a_{4}}^{a_{1}a_{2}}%
\mathcal{R}_{\mu _{1}\mu _{2}}^{a_{3}a_{4}}),  \label{5}
\end{equation}%
where $\xi =\pm 1$ or $\xi =\mp i$, depending whether the signature of $\eta
_{ab}$ is Euclidean or Lorentzian, respectively.

It is worth remarking that (5) determines that $^{\pm }\mathcal{R}_{\mu
_{1}\mu _{2}}^{a_{1}a_{2}}$ is self-dual (antiself-dual) in the sense that%
\begin{equation}
^{\ast \pm }\mathcal{R}_{\mu _{1}\mu _{2}}^{a_{1}a_{2}}\equiv \frac{1}{2}%
\epsilon _{a_{3}a_{4}}^{a_{1}a_{2}}\text{ }^{\pm }\mathcal{R}_{\mu _{1}\mu
_{2}}^{a_{3}a_{4}}=\varepsilon \text{ }\xi \text{ }^{\pm }\mathcal{R}_{\mu
_{1}\mu _{2}}^{a_{1}a_{2}},  \label{6}
\end{equation}%
where, $\varepsilon =+1$ if $\xi =\pm 1$ or $\varepsilon =-1$ if $\xi =\mp i$%
. It is important to note that this result is directly related to the
space-time dimensionality $D$; in this case $D=4$. In fact, this can be
established by the number of indices in $\epsilon _{a_{1}a_{2}a_{3}a_{4}}$
(in this case $4$-indices).

Moreover, considering (5) one finds that (1) can also be written as

\begin{equation}
^{\pm }\mathcal{R}_{\mu _{1}\mu _{2}}^{a_{1}a_{2}}=\text{ }^{\pm }R_{\mu
_{1}\mu _{2}}^{a_{1}a_{2}}+\text{ }^{\pm }\Sigma _{\mu _{1}\mu
_{2}}^{a_{1}a_{2}}.  \label{7}
\end{equation}%
When one substitutes the relation (7) into (4) one obtains

\begin{equation}
S=S_{1}+S_{2}+S_{3},  \label{8}
\end{equation}%
where

\begin{equation}
S_{1}=\frac{1}{2^{2}}\int d^{4}x\epsilon ^{\mu _{1}\mu _{2}\mu _{3}\mu
_{4}}\epsilon _{a_{1}a_{2}a_{3}a_{4}}\text{ }^{\pm }R_{\mu _{1}\mu
_{2}}^{a_{1}a_{2}}\text{ }^{\pm }R_{\mu _{3}\mu _{4}}^{a_{3}a_{4}},
\label{9}
\end{equation}

\begin{equation}
S_{2}=\frac{1}{2}\int d^{4}x\epsilon ^{\mu _{1}\mu _{2}\mu _{3}\mu
_{4}}\epsilon _{a_{1}a_{2}a_{3}a_{4}}\text{ }^{\pm }\Sigma _{\mu _{1}\mu
_{2}}^{a_{1}a_{2}}\text{ }^{\pm }R_{\mu _{3}\mu _{4}}^{a_{3}a_{4}}
\label{10}
\end{equation}%
and

\begin{equation}
S_{3}=\frac{1}{2^{2}}\int d^{4}x\epsilon ^{\mu _{1}\mu _{2}\mu _{3}\mu
_{4}}\epsilon _{a_{1}a_{2}a_{3}a_{4}}\text{ }^{\pm }\Sigma _{\mu _{1}\mu
_{2}}^{a_{1}a_{2}}\text{ }^{\pm }\Sigma _{\mu _{3}\mu _{4}}^{a_{3}a_{4}}.
\label{11}
\end{equation}%
One can prove that $S_{1}$ is related to the Euler and Pontryagin
topological invariants, $S_{2}$ leads to the Einstein-Hilbert action, while $%
S_{3}$ is related to the cosmological constant term (see Ref. [22] for
details).

\bigskip \ 

\noindent \textbf{3.- Ashtekar formalism in 8-dimensions}

\smallskip \ 

In this section, we generalize the procedure of the previous section to $8$%
-dimensions. For this purpose let us first note that the task does not seem
so straightforward since we now need to deal with an $\epsilon $-symbol with 
$8$-indices, namely, $\epsilon _{a_{1}...a_{8}}$. But the curvature tensor $%
R_{\mu _{1}\mu _{2}}^{a_{1}a_{2}}$ contains only the two indices $a_{1}$ and 
$a_{2}$. This is one of the reason why in Refs [6]-[9] an Ashtekar formalism
with an octonionic structure was proposed. Although this an interesting
route here we shall follow an alternative approach. In fact, we shall insist
in using $\epsilon ^{a_{1}a_{2}a_{3}a_{4}a_{5}a_{6}a_{7}a_{8}}$ instead of
the octonionic structure constants. We should mention that just to avoid
additional complication by changing the indices notation, in this section we
shall use the same indices that in the previous section, but now both Latin
and Greek indices shall run from $1$ to $8$.

Consider the tensor

\begin{equation}
\Omega _{\mu _{1}\mu _{2}\mu _{3}\mu _{4}}^{a_{1}a_{2}a_{3}a_{4}}=\delta
_{b_{1}b_{2}b_{3}b_{4}}^{a_{1}a_{2}a_{3}a_{4}}\mathcal{R}_{\mu _{1}\mu
_{2}}^{b_{1}b_{2}}\mathcal{R}_{\mu _{3}\mu _{4}}^{b_{3}b_{4}},  \label{12}
\end{equation}%
where the quantity $\delta _{b_{1}b_{2}b_{3}b_{4}}^{a_{1}a_{2}a_{3}a_{4}}$
is a generalized Kronecker delta. In fact, we shall introduce the definition

\begin{equation}
\epsilon \delta _{b_{1}...b_{8}}^{a_{1}...a_{8}}=\epsilon
^{a_{1}...a_{8}}\epsilon _{b_{1}...b_{8}}.  \label{13}
\end{equation}%
Here, the parameter $\epsilon $ takes the values $\epsilon =+1$ or $\epsilon
=-1$, depending whether the signature of $\eta _{ab}$ is Euclidean of
Lorentzian, respectively. Our new proposed action in $8$-dimensions is%
\begin{equation}
S=\frac{1}{(4!)^{2}}\int d^{8}x\epsilon ^{\mu _{1}....\mu _{8}}\epsilon
_{a_{1}...a_{8}}\text{ }^{\pm }\Omega _{\mu _{1}\mu _{2}\mu _{3}\mu
_{4}}^{a_{1}a_{2}a_{3}a_{4}}\text{ }^{\pm }\Omega _{\mu _{5}\mu _{6}\mu
_{7}\mu _{8}}^{a_{5}a_{6}a_{7}a_{8}}.  \label{14}
\end{equation}%
This is, of course, the analogue in $8$-dimensions of the action (4) in $4$%
-dimensions. Here, one has

\begin{equation}
^{\pm }\Omega _{\mu _{1}\mu _{2}\mu _{3}\mu _{4}}^{a_{1}a_{2}a_{3}a_{4}}=%
\frac{1}{2}(\Omega _{\mu _{1}\mu _{2}\mu _{3}\mu
_{4}}^{a_{1}a_{2}a_{3}a_{4}}+\frac{\xi }{4!}\epsilon
_{a_{5}a_{6}a_{7}a_{8}}^{a_{1}a_{2}a_{3}a_{4}}\Omega _{\mu _{1}\mu _{2}\mu
_{3}\mu _{4}}^{a_{5}a_{6}a_{7}a_{8}}),  \label{15}
\end{equation}%
where $\xi =\pm 1$ or $\xi =\mp i$, depending whether the signature of $\eta
_{ab}$ is Euclidean or Lorentzian, respectively. Using the property (13) of $%
\epsilon _{a_{5}a_{6}a_{7}a_{8}}^{a_{1}a_{2}a_{3}a_{4}}$ one can verify that 
$^{\pm }\Omega _{\mu _{1}\mu _{2}\mu _{3}\mu _{4}}^{a_{1}a_{2}a_{3}a_{4}}$
is in fact self (antiself)-dual quantity, namely

\begin{equation}
^{\ast \pm }\Omega _{\mu _{1}\mu _{2}\mu _{3}\mu
_{4}}^{a_{1}a_{2}a_{3}a_{4}}=\frac{1}{4!}\epsilon
_{a_{5}a_{6}a_{7}a_{8}}^{a_{1}a_{2}a_{3}a_{4}}\text{ }^{\pm }\Omega _{\mu
_{1}\mu _{2}\mu _{3}\mu _{4}}^{a_{5}a_{6}a_{7}a_{8}}=\varepsilon \xi \text{ }%
^{\pm }\Omega _{\mu _{1}\mu _{2}\mu _{3}\mu _{4}}^{a_{1}a_{2}a_{3}a_{4}},
\label{16}
\end{equation}%
where, $\varepsilon =+1$ if $\xi =\pm 1$ or $\varepsilon =-1$ if $\xi =\mp i$%
. This means that one of the key properties of self (antiself)-duality in
the Ashtekar formalism in $4$-dimensions is preserved in the action (14) in $%
8$-dimensions. Note that since $\xi ^{2}=\pm $ one obtains%
\begin{equation}
^{\ast \ast \pm }\Omega _{\mu _{1}\mu _{2}\mu _{3}\mu
_{4}}^{a_{1}a_{2}a_{3}a_{4}}=\pm \text{ }^{\pm }\Omega _{\mu _{1}\mu _{2}\mu
_{3}\mu _{4}}^{a_{1}a_{2}a_{3}a_{4}}.  \label{17}
\end{equation}%
The dual of the dual of an object is equal to the same object, that is, one
has the dual property $^{\ast \ast }=\pm I$.

\bigskip \ 

\noindent \textbf{4.- Reduction of the Ashtekar formalism in 8-dimensions}

\smallskip \ 

In this section, we develop some of the consequences of the action (14).
Substituting (15) into (14) leads to

\begin{equation}
\begin{array}{c}
S=\frac{1}{4(4!)^{2}}\int d^{8}x\epsilon ^{\mu _{1}....\mu _{8}}\epsilon
_{a_{1}...a_{8}}(\Omega _{\mu _{1}\mu _{2}\mu _{3}\mu
_{4}}^{a_{1}a_{2}a_{3}a_{4}}+\frac{\xi }{4!}\epsilon
_{b_{1}b_{2}b_{3}b_{4}}^{a_{1}a_{2}a_{3}a_{4}}\Omega _{\mu _{1}\mu _{2}\mu
_{3}\mu _{4}}^{b_{1}b_{2}b_{3}b_{4}}) \\ 
\\ 
\times (\Omega _{\mu _{5}\mu _{6}\mu _{7}\mu _{8}}^{a_{5}a_{6}a_{7}a_{8}}+%
\frac{\xi }{4!}\epsilon _{c_{1}c_{2}c_{3}c_{4}}^{a_{5}a_{6}a_{7}a_{8}}\Omega
_{\mu _{5}\mu _{6}\mu _{7}\mu _{8}}^{c_{1}c_{2}c_{3}c_{4}}).%
\end{array}
\label{18}
\end{equation}%
Simplifying this expression one gets

\begin{equation}
\begin{array}{c}
S=\frac{1}{2(4!)^{2}}\int d^{8}x\epsilon ^{\mu _{1}....\mu _{8}}\epsilon
_{a_{1}...a_{8}}\Omega _{\mu _{1}\mu _{2}\mu _{3}\mu
_{4}}^{a_{1}a_{2}a_{3}a_{4}}\Omega _{\mu _{5}\mu _{6}\mu _{7}\mu
_{8}}^{a_{5}a_{6}a_{7}a_{8}} \\ 
\\ 
+\frac{\xi }{2(4!)}\int d^{8}x\epsilon ^{\mu _{1}....\mu _{8}}\eta
_{a_{1}a_{5}}\eta _{a_{2}a_{6}}\eta _{a_{3}a_{7}}\eta _{a_{4}a_{8}}\Omega
_{\mu _{1}\mu _{2}\mu _{3}\mu _{4}}^{a_{1}a_{2}a_{3}a_{4}}\Omega _{\mu
_{5}\mu _{6}\mu _{7}\mu _{8}}^{a_{5}a_{6}a_{7}a_{8}}.%
\end{array}
\label{19}
\end{equation}%
Using (12) this expression becomes

\begin{equation}
\begin{array}{c}
S=\frac{1}{2}\int d^{8}x\epsilon ^{\mu _{1}....\mu _{8}}\epsilon
_{a_{1}...a_{8}}\mathcal{R}_{\mu _{1}\mu _{2}}^{a_{1}a_{2}}\mathcal{R}_{\mu
_{3}\mu _{4}}^{a_{3}a_{4}}\mathcal{R}_{\mu _{5}\mu _{6}}^{a_{5}a_{6}}%
\mathcal{R}_{\mu _{7}\mu _{8}}^{a_{7}a_{8}} \\ 
\\ 
+\frac{\epsilon }{2}\xi \int d^{8}x\epsilon ^{\mu _{1}....\mu _{8}}\eta
_{a_{1}..a_{4},a_{5}...a_{8}}\mathcal{R}_{\mu _{1}\mu _{2}}^{a_{1}a_{2}}%
\mathcal{R}_{\mu _{3}\mu _{4}}^{a_{3}a_{4}}\mathcal{R}_{\mu _{5}\mu
_{6}}^{a_{5}a_{6}}\mathcal{R}_{\mu _{7}\mu _{8}}^{a_{7}a_{8}},%
\end{array}
\label{20}
\end{equation}%
where

\begin{equation}
\eta _{a_{1}..a_{4},a_{5}...a_{8}}=\delta
_{a_{5}a_{6}a_{7}a_{8}}^{b_{1}b_{2}b_{3}b_{4}}\eta _{a_{1}b_{1}}\eta
_{a_{2}b_{2}}\eta _{a_{3}b_{3}}\eta _{a_{4}b_{4}}.  \label{21}
\end{equation}%
Now, substituting (1) into (20) with $\lambda =1$ one obtains the action

\begin{equation}
S=S_{1}+S_{2}+S_{3}+S_{4}+S_{5}+S_{6}+S_{7}+S_{8}+S_{9}+S_{10},  \label{22}
\end{equation}%
where

\begin{equation}
S_{1}=\frac{1}{2}\int d^{8}x\epsilon ^{\mu _{1}....\mu _{8}}\epsilon
_{a_{1}...a_{8}}R_{\mu _{1}\mu _{2}}^{a_{1}a_{2}}R_{\mu _{3}\mu
_{4}}^{a_{3}a_{4}}R_{\mu _{5}\mu _{6}}^{a_{5}a_{6}}R_{\mu _{7}\mu
_{8}}^{a_{7}a_{8}},  \label{23}
\end{equation}

\begin{equation}
S_{2}=2\int d^{8}x\epsilon ^{\mu _{1}....\mu _{8}}\epsilon
_{a_{1}...a_{8}}\Sigma _{\mu _{1}\mu _{2}}^{a_{1}a_{2}}R_{\mu _{3}\mu
_{4}}^{a_{3}a_{4}}R_{\mu _{5}\mu _{6}}^{a_{5}a_{6}}R_{\mu _{7}\mu
_{8}}^{a_{7}a_{8}},  \label{24}
\end{equation}%
\begin{equation}
S_{3}=3\int d^{8}x\epsilon ^{\mu _{1}....\mu _{8}}\epsilon
_{a_{1}...a_{8}}\Sigma _{\mu _{1}\mu _{2}}^{a_{1}a_{2}}\Sigma _{\mu _{3}\mu
_{4}}^{a_{3}a_{4}}R_{\mu _{5}\mu _{6}}^{a_{5}a_{6}}R_{\mu _{7}\mu
_{8}}^{a_{7}a_{8}},  \label{25}
\end{equation}%
\begin{equation}
S_{4}=2\int d^{8}x\epsilon ^{\mu _{1}....\mu _{8}}\epsilon
_{a_{1}...a_{8}}\Sigma _{\mu _{1}\mu _{2}}^{a_{1}a_{2}}\Sigma _{\mu _{3}\mu
_{4}}^{a_{3}a_{4}}\Sigma _{\mu _{5}\mu _{6}}^{a_{5}a_{6}}R_{\mu _{7}\mu
_{8}}^{a_{7}a_{8}},  \label{26}
\end{equation}

\begin{equation}
S_{5}=\frac{1}{2}\int d^{8}x\epsilon ^{\mu _{1}....\mu _{8}}\epsilon
_{a_{1}...a_{8}}\Sigma _{\mu _{1}\mu _{2}}^{a_{1}a_{2}}\Sigma _{\mu _{3}\mu
_{4}}^{a_{3}a_{4}}\Sigma _{\mu _{5}\mu _{6}}^{a_{5}a_{6}}\Sigma _{\mu
_{7}\mu _{8}}^{a_{7}a_{8}},  \label{27}
\end{equation}%
and%
\begin{equation}
S_{6}=\frac{\epsilon }{2}\xi \int d^{8}x\epsilon ^{\mu _{1}....\mu _{8}}\eta
_{a_{1}..a_{4},a_{5}...a_{8}}R_{\mu _{1}\mu _{2}}^{a_{1}a_{2}}R_{\mu _{3}\mu
_{4}}^{a_{3}a_{4}}R_{\mu _{5}\mu _{6}}^{a_{5}a_{6}}R_{\mu _{7}\mu
_{8}}^{a_{7}a_{8}},  \label{28}
\end{equation}

\begin{equation}
S_{7}=2\epsilon \xi \int d^{8}x\epsilon ^{\mu _{1}....\mu _{8}}\eta
_{a_{1}..a_{4},a_{5}...a_{8}}\Sigma _{\mu _{1}\mu _{2}}^{a_{1}a_{2}}R_{\mu
_{1}\mu _{2}}^{a_{3}a_{4}}R_{\mu _{5}\mu _{6}}^{a_{5}a_{6}}R_{\mu _{7}\mu
_{8}}^{a_{7}a_{8}},  \label{29}
\end{equation}%
\begin{equation}
S_{8}=3\epsilon \xi \int d^{8}x\epsilon ^{\mu _{1}....\mu _{8}}\eta
_{a_{1}..a_{4},a_{5}...a_{8}}\Sigma _{\mu _{1}\mu _{2}}^{a_{1}a_{2}}\Sigma
_{\mu _{3}\mu _{4}}^{a_{3}a_{4}}R_{\mu _{5}\mu _{6}}^{a_{5}a_{6}}R_{\mu
_{7}\mu _{8}}^{a_{7}a_{8}},  \label{30}
\end{equation}%
\begin{equation}
S_{9}=2\epsilon \xi \int d^{8}x\epsilon ^{\mu _{1}....\mu _{8}}\eta
_{a_{1}..a_{4},a_{5}...a_{8}}\Sigma _{\mu _{1}\mu _{2}}^{a_{1}a_{2}}\Sigma
_{\mu _{3}\mu _{4}}^{a_{3}ba}\Sigma _{\mu _{5}\mu _{6}}^{a_{5}a_{6}}R_{\mu
_{7}\mu _{8}}^{a_{7}a_{8}},  \label{31}
\end{equation}

\begin{equation}
S_{10}=\frac{\epsilon }{2}\xi \int d^{8}x\epsilon ^{\mu _{1}....\mu
_{8}}\eta _{a_{1}..a_{4},a_{5}...a_{8}}\Sigma _{\mu _{1}\mu
_{2}}^{a_{1}a_{2}}\Sigma _{\mu _{3}\mu _{4}}^{a_{3}a_{4}}\Sigma _{\mu
_{5}\mu _{6}}^{a_{5}a_{6}}\Sigma _{\mu _{7}\mu _{8}}^{a_{7}a_{8}}.
\label{32}
\end{equation}

Using the antisymmetry/symmetry properties of $\epsilon ^{\mu _{1}....\mu
_{8}}$ and $R_{\mu _{1}\mu _{2}\mu _{5}\mu _{6}}=e_{\mu _{1}a_{1}}e_{\mu
_{2}a_{2}}R_{\mu _{5}\mu _{6}}^{a_{1}a_{2}}$ it is not difficult to verify
that

\begin{equation}
S_{7}=0,S_{8}=0,S_{9}=0,S_{10}=0.  \label{33}
\end{equation}%
Thus, (22) is reduced to

\begin{equation}
S=S_{1}+S_{2}+S_{3}+S_{4}+S_{5}+S_{6}.  \label{34}
\end{equation}%
It turns out that $S_{1}$ and $S_{6}$ can be identified with a Euler and
Pontryagin topological invariant terms in $8$-dimensions. While, $S_{5}$ is
a cosmological constant term and $S_{4}$ leads to the Einstein-Hilbert
action in $8$-dimensions. Finally, through a long computation one finds that 
$S_{2}$ and $S_{3}$ become%
\begin{equation}
\begin{array}{c}
S_{2}=2^{6}\int d^{8}x\sqrt{\epsilon g}[R^{3}-12RR_{\mu \nu }R^{\mu \nu
}+3RR_{\mu \nu \alpha \beta }R^{\mu \nu \alpha \beta } \\ 
\\ 
+16R_{\mu \nu }R_{\alpha }^{\mu }R^{\nu \alpha }+24R_{\mu \nu }R_{\alpha
\beta }R^{\mu \nu \alpha \beta }-24R_{\mu \nu }R_{\alpha \beta \lambda
}^{\mu }R^{\nu \alpha \beta \lambda } \\ 
\\ 
+2R_{\alpha \beta }^{\mu \nu }R_{\tau \lambda }^{\alpha \beta }R_{\mu \nu
}^{\tau \lambda }-8R_{\alpha \beta }^{\mu \nu }R_{\nu \lambda }^{\beta \tau
}R_{\tau \mu }^{\lambda \alpha }]%
\end{array}
\label{35}
\end{equation}%
and%
\begin{equation}
S_{3}=3(2)^{4}(4!)\int d^{8}x\sqrt{\epsilon g}[R^{2}-4R^{\mu \nu }R_{\mu \nu
}+R^{\mu \nu \alpha \beta }R_{\mu \nu \alpha \beta }],  \label{36}
\end{equation}%
respectively. Here, $R_{\mu \nu \alpha \beta }=e_{a\alpha }e_{b\beta }R_{\mu
\nu }^{ab}$, $R_{\mu \nu }=R_{\mu \alpha \nu }^{\alpha }$ and $R=g^{\mu \nu
}R_{\mu \nu }$. Surprisingly, one finds that (35) and (36) have exactly the
same form that if one considers the Lovelock theory [23] (see Appendix A en
Ref. [24]). In fact, since $S_{7},S_{8},S_{9},$ and $S_{10}$ vanishes one
discovers that the sum of $S_{1},S_{2},S_{3},S_{4},S_{5}$ and $S_{6}$
describe a Lovelock action in $8$-dimensions. Moreover, since $S_{1}$ and $%
S_{6}$ are topological invariant terms one sees that the whole dynamics of
the system is contained in $S_{2},S_{3},S_{4}$ and $S_{5}$.

\smallskip \ 

\noindent \textbf{5.- Final remarks}

\smallskip \ 

Summarizing, in this article, we have proposed the action (14) as an
alternative description of the Ashtekar formalism in $8$-dimensions and, in
section 4, we have shown some of its consequences. In particular we have
shown that it reduces to the Lovelock action in $8$-dimensions. It remains
to analyze the implications of (14) at the quantum level.

Thinking in terms of division algebras one notes that the $4$-dimensional
and $8$-dimensional structures are two of the allowed dimensions in such
algebras. The other are $1$-dimensional and $2$-dimensional. So the original 
$4$-dimensional Ashtekar formalism corresponds to one of these division
algebras. Thus, from the division algebras point of view the $8$-dimensional
Ashtekar formalism must be considered equally important. Of course, $1,2,4$
and $8$ dimensions are closely related with the real, complex, quaternion
and octonionic numbers via the Hurwitz theorem (see Ref. [25] and references
therein). So, it may be interesting to see whether both proposals of the
Ashtekar formalism in $8$-dimensions, the one based on an octonionic
structure and one presented here, are related. A direct suggestion for this
link is provided by the self (antiself)-duality relation

\begin{equation}
\eta ^{a_{1}a_{2}a_{3}a_{4}}=\pm \epsilon
^{a_{1}a_{2}a_{3}a_{4}a_{5}a_{6}a_{7}a_{8}}\eta _{a_{5}a_{6}a_{7}a_{8}},
\label{37}
\end{equation}%
between the octonionic structure constants $\eta ^{a_{1}a_{2}a_{3}a_{4}}$
and the $\epsilon $-symbol in $8$-dimensions $\epsilon
^{a_{1}a_{2}a_{3}a_{4}a_{5}a_{6}a_{7}a_{8}}$. Another possibility for such a
link comes from group analysis in the sense that both theories are connected
with the group $SO(8)$ which can be broken as $SO(8)\rightarrow S_{7}\times
S_{7}\times G_{2}$, where $S_{7}$ is the parallelizable seven sphere and $%
G_{2}$ is one of the exceptional groups. As it is known the
parallelizability of $S_{7}$ is related with the existence of the octonionic
structure in $8$-dimensions.

We have developed a procedure which can be used in any signature specified
by the flat metric $\eta _{ab}$ in $8$-dimensions. So, in principle, one may
think in the particular case of the $(4+4)$-signature which in turn can be
associated with the $M$-theory proposal of $(5+5)$-dimensions. In this
sense, it may be interesting for further research to see whether the present
self (antiself)-dual Ashtekar formalism in $8$-dimensions can be connected
with the oriented matroid theory which has been proposed as a mathematical
framework for $M$-theory (see Refs. [17]-[19] and references therein).

Moreover, in $(1+3)$-dimensions, the MacDowell-Mansouri formalism is based
on the de Sitter gauge group $SO(1,4)$ (or anti-de Sitter gauge group $%
SO(2,3)$). In fact, at the level of the spin connection $\omega _{\mu
}^{AB}=(e_{\mu }^{a},\omega _{\mu }^{ab})$, in such a formalism, one
considers the transition $SO(1,4)\rightarrow SO(1,3)$ (or $%
SO(2,3)\rightarrow SO(1,3)$) obtaining the curvature $\mathcal{R}_{\mu \nu
}^{ab}$ as expressed in (1)-(3). In $(1+7)$-dimensions, one would expect
that the curvature $\mathcal{R}_{\mu \nu }^{ab}$ considered in the proposed
action (14) is the result of the transition $SO(1,8)\rightarrow SO(1,7)$ (or 
$SO(2,7)\rightarrow SO(1,7)$). In both cases, in $(1+3)$-dimensions and $%
(1+7)$-dimensions, such a transition is reflected in the cosmological
constant term. In some sense, one can say that the original symmetries $%
SO(1,4)$ (or $SO(2,3)$) in the case of $(1+3)$-dimensions and $SO(1,8)$ (or $%
SO(2,7)$) in the case of $(1+7)$-dimensions are hidden symmetries of the
reduced actions (4) and (14), respectively. Thus assuming a non-vanishing
cosmological constant one can focus in the gauge symmetry group $SO(1,3)$ in
the case of $(1+3)$-dimensions and $SO(1,7)$ in the case of $(1+7)$%
-dimensions. These comments can be clarified further recalling that in $%
(1+3) $-dimensions the algebra $so(1,3)$ can be written as $%
so(1,3)=su(2)\times su(2)$. So, the curvature $\mathcal{R}^{ab}$ can be
decomposed additively [2]: $\mathcal{R}^{ab}(\omega )=\quad ^{+}\mathcal{R}%
^{ab}(^{+}\omega )+^{-}\mathcal{R}^{ab}(^{-}\omega )$ where $^{+}\omega $
and $^{-}\omega $ are the self-dual and anti-self-dual parts of the spin
connection $\omega $. In an Euclidean context, this is equivalent of writing
the norm group of quaternions $SO(4)$ as $SO(4)=S^{3}\times S^{3}$, where $%
S^{3}$ denotes the three sphere. The situation in eight dimensions is very
similar since $SO(8)=S^{7}\times S^{7}\times G_{2}$, with $S^{7}$ denoting
the seven sphere, suggesting that one can also define duality in eight
dimensions, but modulo the exceptional group $G_{2}$ [26]-[27]. At the level
of $1+7$-dimensions, the situation is not so simple since the closest
decomposition to the case $so(1,3)$ of the Lie algebra $so(1,7)$ associated
with $SO(1,7)$ seems to be $so(1,7)=g_{2}\oplus L_{\func{Im}(\mathcal{O}%
)}\oplus R_{\func{Im}(\mathcal{O})}$, where $g_{2}$ is the Lie algebra of
the exceptional group $G_{2}$ and $L_{\func{Im}(\mathcal{O})}$ ($R_{\func{Im}%
(\mathcal{O})})$ is the space of linear transformations of $\mathcal{O}$
given by left (right) multiplication by imaginary octonions. A possible
check that this is really the case observe that $\dim so(1,7)=28$ and $\dim
g_{2}+\dim L_{\func{Im}(\mathcal{O})}+\dim R_{\func{Im}(\mathcal{O})}=14+7+7$
(see Ref. [26] for details)$.$ Thus while the self-dual sector is related to
the exceptional group $G_{2}$ one finds the intriguing result that the
antiself-dual sector should be related to two copies of $\func{Im}(\mathcal{O%
})$, the $7-$dimensional space consisting of all imaginary octonions.

In the case of $(4+4)$-dimensions one may consider the chain of maximal
embeddings and branches,%
\begin{equation}
so(4,4)\supset s(2,R)\oplus so(2,3)\supset so(1,1)\oplus sl(2,R)\oplus
sl(2,2).  \label{38}
\end{equation}%
However, these subgroups are not full symmetry groups and therefore it is
difficult to reveal hidden symmetries in the action (14).

Finally, one may also think in a generalization of our procedure to other
dimensions beyond $4$ and $8$ dimensions. Let us consider a generalization
of (12) in the form

\begin{equation}
\Omega _{\mu _{1}\mu _{2}...\mu _{D/2-1}\mu
_{D/2}}^{a_{1}a_{2}...a_{D/2-1}a_{D/2}}=\delta
_{b_{1}b_{2}...b_{D/2-1}b_{D/2}}^{a_{1}a_{2}...a_{D/2-1}a_{D/2}}\mathcal{R}%
_{\mu _{1}\mu _{2}}^{b_{1}b_{2}}...\mathcal{R}_{\mu _{D/2-1}\mu
_{D/2}}^{b_{D/2-1}b_{D/2}}.  \label{39}
\end{equation}%
One observes that due to the fact that the curvature $\mathcal{R}_{\mu \nu
}^{ab}$ is a two form the formalism is possible only in $2^{2l}$-dimensions
(or $2^{2l-1}$-dimensions), where $l$ denotes the number odd (or even) $%
\mathcal{R}$-terms in (39). Thus, the generalized action (14) can be written
as

\begin{equation}
S=\frac{1}{A}\int d^{D}x\epsilon ^{\mu _{1}\mu _{2}....\mu _{D}}\epsilon
_{a_{1}a_{2}...a_{D}}\text{ }^{\pm }\Omega _{\mu _{1}\mu _{2}...\mu
_{D/2}}^{a_{1}a_{2}...a_{D/2}}{}^{\pm }\Omega _{\mu _{D/2+1}\mu
_{D/2+2}...\mu _{D}}^{a_{D/2+1}a_{D/2+2}...a_{D}},  \label{40}
\end{equation}%
where $D=2^{2l}$ or $D=2^{2l-1}$ and $A$ is a proper numerical factor. Here, 
$^{\pm }\Omega _{\mu _{1}\mu _{2}...\mu _{D/2}}^{a_{1}a_{2}...a_{D/2}}$ is
defined as in (15) but using $\epsilon _{a_{1}...a_{D}}$. Of course, one may
expect that, up to topological invariant terms, an expansion of (40) in
terms of $\mathcal{R}_{\mu \nu }^{ab}$ shall lead to an action with various
higher dimensional Lovelock type terms. For further research, it may be
interesting to explore this observation in more detail.

\bigskip \ 

\begin{center}
\textbf{Acknowledgments}

\smallskip \ 
\end{center}

I would like to thank the referee for his valuable remarks. Also, I would
like to thank P. A. Nieto for his helpful comments. This work was partially
supported by PROFAPI-UAS/2012.

\bigskip \


\begin{thebibliography}{99}
\bibitem{1} N. Bodendorfer, T. Thiemann, and A. Thurn, Class.Quant.Grav. 
\textbf{30} (2013) 045001; arXiv:1105.3703 [gr-qc].

\bibitem{2} N. Bodendorfer, T. Thiemann, and A. Thurn, Class.Quant.Grav. 
\textbf{30} (2013) 045002; arXiv:1105.3704 [gr-qc].

\bibitem{3} N. Bodendorfer, T. Thiemann, and A. Thurn, Class.Quant.Grav. 
\textbf{30} (2013) 045003; arXiv:1105.3705 [gr-qc].

\bibitem{4} N. Bodendorfer, T. Thiemann, and A. Thurn, Class.Quant.Grav. 
\textbf{30} (2013) 045004; arXiv:1105.3706 [gr-qc].

\bibitem{5} N. Bodendorfer, T. Thiemann, and A. Thurn, Class.Quant.Grav. 
\textbf{31} (2014) 055002; arXiv:1304.2679 [gr-qc].

\bibitem{6} J. A. Nieto, Class. Quant. Grav. \textbf{22} (2005) 947-955;
hep-th/0410260.

\bibitem{7} J. A. Nieto, Mod. Phys. Lett. A \textbf{20} (2005) 2157-2163;
hep-th/0411124.

\bibitem{8} J. A. Nieto, Gen. Rel. Grav. \textbf{39} (2007) 1109;
hep-th/0506253.

\bibitem{9} J. A. Nieto, \textquotedblleft Towards a background independent
quantum gravity in eight dimensions\textquotedblright , arXiv:0704.2769
[hep-th].

\bibitem{10} J. A. Nieto, Rev. Mex. Fis. \textbf{57} (2011) 400;
arXiv:1003.4750 [hep-th].

\bibitem{11} M. Green, J. Schwarz, E. Witten, \textit{Superstring Theory}
(Cambridge U. Press, Cambridge,UK, 1987).

\bibitem{12} C. Hull and R. R. Khuri, Nucl. Phys. B \textbf{536} (1998) 219;
e-Print: hep-th/9808069.

\bibitem{13} M. A. De Andrade, M. Rojas and F. Toppan, Int. J. Mod. Phys. A 
\textbf{16} (2001) 4453; hep-th/0005035.

\bibitem{14} M. Rojas, M. A. De Andrade, L. P. Colatto, J. L. Matheus-Valle,
L. P. G. De Assis and J. A. Helayel-Neto, \textquotedblleft Mass Generation
and Related Issues from Exotic Higher Dimensions\textquotedblright : arXiv:
1111.2261 [hep-th].

\bibitem{15} M. A. De Andrade, I. V. Vancea., \textquotedblleft Action for
spinor fields in arbitrary dimensions\textquotedblright , e-Print:
hep-th/0105025.

\bibitem{16} P. van Nieuwenhuizen, \textit{Supergravity}, Phys. Rep. \textbf{%
68}, (1981) 189.

\bibitem{17} J. A. Nieto, Nucl. Phys. B \textbf{883} (2014) 350;
arXiv:1402.6998 [hep-th].

\bibitem{18} J. A. Nieto, Adv. Theor. Math. Phys. \textbf{8} (2004) 177;\
hep-th/0310071.

\bibitem{19} J .A. Nieto, Adv. Theor. Math. Phys. \textbf{10} (2006) 747;
hep-th/0506106.

\bibitem{20} J. A. Nieto, Phys. Lett. B \textbf{718} (2013) 1543;
arXiv:1210.0928 [hep-th] .

\bibitem{21} J. A. Nieto, Phys. Lett. B \textbf{692} (2010) 43;
arXiv:1004.5372 [hep-th].

\bibitem{22} J. A. Nieto, O. Obregon and J. Socorro, Phys. Rev. D \textbf{50}
(1994) 3583; gr-qc/9402029.

\bibitem{23} D. J. Lovelock, Math. Phys. \textbf{12} (1971) 498.

\bibitem{24} M. Cruz and E. Rojas, Class. Quant. Grav. \textbf{30} (2013)
115012; arXiv:1212.1704.

\bibitem{25} J. A. Nieto, J. Mod. Phys. \textbf{4} (2013) 32;
arXiv:1303.2673 [hep-th].

\bibitem{26} A. R. Dundarer, F. Gursey and C. H. Tze, J. Math. Phys. \textbf{%
25} (1984) 1496.

\bibitem{27} A. R. Dundarer and F. Gursey, J. Math. Phys. \textbf{32},
(1991) 1178.
\end{thebibliography}
\end{document}